\documentclass{article}

\usepackage{amsfonts}
\usepackage{amssymb}
\usepackage[mathscr]{eucal}
\usepackage{url}
\usepackage{alg}
\usepackage{graphicx}

\newcommand{\su}{\phantom{\mbox{$\overline{{\overline{\mid}}}$}}}
\newcommand{\giu}{\phantom{\mbox{${\underline{\underline{\mid}}}$}}}

\newcommand{\tab}{\hspace*{0.314cm}}

\newtheorem{example}{Example}

\newcommand{\name}{\mbox{$\mathcal{B}^{\mbox{\sf\tiny AAC}}$}}
\newcommand{\BAAC}{\name}
\newcommand{\BMV}{\mbox{${B}^{\mbox{\sf\tiny MV}}$}}
\newcommand{\BMAP}{\mbox{${B}^{\mbox{\sf\tiny MAP}}$}}
\newcommand{\codetext}[1]{\mbox{\tt #1}}
\newcommand{\nat}{\mathbb{N}}
\newcommand{\interi}{\mathbb{Z}}
\newcommand{\op}{\mbox{\codetext{op}\hspace*{0.5ex}}}
\newcommand{\N}{\mathbf{N}}

\newcommand{\eoe}{ ~~\hspace*{\fill}~\mbox{$\Box$}}
\newcommand{\spv}{1.5ex}
\newcommand{\Prec}{Prec}
\newcommand{\n}{\mathbf{N}}
\newcommand{\FV}{\mathsf{fluents}}
\newcommand{\ine}{\codetext{ine}}
\newcommand{\Eff}{\codetext{Eff}}
\newcommand{\DEff}{\codetext{DEff}}

\newcommand{\back}{\mbox{$^{-1}$}}

\title{{Autonomous Agents Coordination: Action  Languages meet CLP($\mathcal{FD}$) and
Linda}\thanks{Research partially funded by
GNCS-INdAM projects,
MUR-PRIN: \emph{Innovative and \mbox{multidisciplinary} approaches
for constraint and preference reasoning} project;
NSF grants IIS-0812267 and HRD-0420407; and grants
2009.010.0336 and 2010.011.0403.
}}

\author
{AGOSTINO DOVIER\\
Universit{\`a} di Udine, 
Dipartimento di Matematica e Informatica\\
{{\tt dovier@dimi.uniud.it}}
\and
ANDREA FORMISANO\\
Universit{\`a} di Perugia, 
Dipartimento di Matematica e Informatica\\
{{\tt formis@dmi.unipg.it}}
\and
ENRICO PONTELLI\\
New Mexico State University, 
 Department of Computer Science\\
{{\tt epontell@cs.nmsu.edu}}
}

\begin{document}

\maketitle

\begin{abstract}
The paper presents a knowledge representation formalism, in the form
of a high-level \emph{Action Description Language (ADL)} for multi-agent systems,
where  autonomous agents reason and act in a shared environment.
Agents are autonomously pursuing
individual goals, but are capable of interacting through a shared knowledge
repository.
In their interactions through shared portions of the world, the
agents deal with problems of synchronization and concurrency; the
action language allows the description of
strategies  to ensure a consistent
 global execution of the agents' autonomously derived plans.
A distributed planning problem is formalized by providing
the declarative specifications of the portion of the problem
pertaining a single agent.
Each of these specifications
is executable by a stand-alone CLP-based planner.
The coordination among agents exploits a
Linda infrastructure.
The proposal is validated in a prototype implementation
developed in
SICStus Prolog.

{\em To appear in
Theory and Practice of Logic Programming (TPLP).}
\end{abstract}

\section{Introduction}\label{sec:intro}

Representing and reasoning in multi-agent domains are two of the most
active research areas in \emph{multi-agent system (MAS)} research. The literature in
this area is extensive, and it provides a plethora of logics for representing and
reasoning about various aspects of MAS domains, e.g.,~\cite{centralized2,Gerbrandy06,HoekJW05,SpaanGV06,fagin95}.

A large number  of the logics proposed in the
literature have been designed to specifically focus on particular aspects of
the problem of modeling MAS,
often justified by a specific application scenario. This makes them suitable
to address specific subsets of the general features required to model
real-world MAS domains.
The task of generalizing some of these existing proposals to create a
uniform and  comprehensive framework for modeling several different aspects
of MAS domains is an open problem. Although we do not dispute the possibility
of extending several of these existing proposals in various directions, the
task does not seem easy. Similarly, a variety of multi-agent programming
platforms have been proposed, mostly in the style of multi-agent programming
languages,
like
Jason~\cite{jason}, ConGolog~\cite{congolog},
3APL~\cite{3apl}, GOAL~\cite{goal}, but with limited planning
capabilities.

 Our effort in this paper  is focused on the development
of a novel action language  for
multi-agent systems.
The foundations of this effort can be found in the
 action language $B^{MV}$~\cite{DFP-TPLP}; this is a flexible
single-agent action language, which  generalizes the
action language $B$ \cite{GL98} with support for multi-valued
fluents, non-Markovian domains, and constraint-based formulations---enabling,
 for example, the formulation of costs and preferences.
$B^{MV}$ has been implemented in CLP($\mathcal{FD}$).

In this
work, we  extend $B^{MV}$ to support MAS domains. The
perspective is that of a distributed environment, with agents pursuing
individual goals but capable of interacting through shared knowledge
and through collaborative actions.
A first step in this direction
has been described in the $B^{\sf MAP}$ language~\cite{DFP-POTSDAM},
a multi-agent action language with capabilities for \emph{centralized}
planning. In this paper, we expand on this by moving $B^{\sf MAP}$ towards
 a truly distributed multi-agent platform. The language is extended
with \emph{C}ommunication
 primitives for modeling interactions among \emph{A}utonomous \emph{A}gents.
 We refer to this language simply as \name.
 Differently from $B^{\sf MAP}$, agents
 in the framework proposed in this paper  have private goals and are capable of developing
 independent plans. Agents' plans are composed in a distributed fashion, leading
 to replanning and/or introduction of communication activities to enable a consistent
 global execution.

 The design of \name\ is validated in a prototype, available from
 \url{http://www.dimi.uniud.it/dovier/BAAC},
 that uses CLP($\mathcal{FD})$ for the
  development of the individual plans of each agent and Linda for the
  coordination and interaction among them.

\section{Syntax of the Multiagent Language \name}\label{sec:suntax}

The signature of \name\ consists of:
\begin{enumerate}
\item  A set
$\mathcal{G}$ of \emph{agent} names, used to identify the agents
in the system;
\item  A set $\mathcal{A}$ of \emph{action} names;
\item  A set $\mathcal{F}$ of \emph{fluent} names---i.e., predicates
describing properties of  objects in the world, and providing  description of
states of the world; such properties might be affected by the execution of actions; and
\item  A set $\mathcal{V}$ of values for the fluents in $\mathcal{F}$---we assume $\mathcal{V}=\interi$.
\end{enumerate}
The behavior of each agent~$a$ is specified by  an action description
theory $\mathcal{D}_a$, composed of  axioms of the forms described next.

Considering the action theory $\mathcal{D}_a$ of an agent $a$, name and priority
of the agent are specified by
\emph{agent declarations}:
\begin{equation}
\label{agentDef}\begin{array}{l}
\codetext{agent } a \codetext{ [ priority } n \codetext{ ]}
\end{array}
\end{equation}
where $n \in \nat$.
We adopt the convention that $0$ denotes the highest priority---which
also represents the default priority  in absence of a
 declaration.
As we will see, priorities can be used to resolve possible conflicts among
actions of different agents.


It is possible to specify which agents are known to the agent $a$, as follows:
\begin{equation}\label{ax:agentknown}
\codetext{known\_agents } a_1, a_2, \dots, a_k
\end{equation}
Agent $a$ can  explicitly communicate with any of the
agents $a_i$, as discussed below.


We assume the existence of a ``global''  set $\mathcal{F}$
of fluents, and  any
agent $a$ knows and can access only those fluents that are
declared in $\mathcal{D}_a$ by axioms of the form:
\begin{equation}\label{fluentDef}
\codetext{fluent } f_1,\ldots,f_h \codetext{ valued } \mathit{dom}_i
\end{equation}
with  $\{f_1, \dots, f_h\}\subseteq \mathcal{F}$,  $h\geq 1$, and $\textit{dom}_i \subset \mathcal{V}$
is a set of values representing
the admissible values for each $f_i$
(possibly represented as an interval $[v_1,v_2]$). These fluents describe
the ``local state'' of the agent.
We assume that the fluents accessed by multiple agents are defined consistently in each
agent's local theory.

\begin{example}\label{volley1}
Let us specify a domain inspired by volleyball.
There are two teams: \emph{black} and \emph{white}, with one
player in each team;
let us focus on the domain for the white team
(Sect.~\ref{volleysect} deals with the case that involves more players).
We introduce fluents to model the positions of the players and
of the ball, the possession of the ball, the score, and a numerical fluent
\codetext{defense\_time}.
All players know the positions of all players.
Since the teams are separated by the net, the
\codetext{x}-coordinates of a black and white players must differ.
This can be stated by:

\codetext{\footnotesize
\begin{tabular}{l}
agent player(white,X)               :- num(X).\su\\
known\_agents player(black,X)  :- num(X).\giu \\
fluent x(player(white,X)) valued [B,E] :-
       num(X), net(NET),B is NET+1, linex(E).\\
fluent x(player(black,X)) valued [1,E] :-
       num(X), net(NET),E is NET-1.\\
fluent y(A) valued [1,MY] :-  player(A), liney(MY).
\\
fluent x(ball) valued [1,MX] :-  linex(MX).\\
fluent y(ball) valued [1,MY] :-  liney(MY).\\
fluent hasball(A) valued [0,1] :- agent(A).\\
fluent point(T) valued [0,1] :- team(T). \\
fluent defense\_time valued [0,1]. \\
team(black). \hspace{.1cm} team(white).  \hspace{.1cm} num(1). \hspace{.1cm}
linex(11).   \hspace{.1cm} net(6).    \hspace{.1cm} liney(5).
\giu
\end{tabular}
}
\par\noindent
where \codetext{linex}/\codetext{liney} are the field sizes, and
\codetext{net} is the \codetext{x}-coordinate of the net.\eoe
\end{example}

\noindent
{F}luents  are used in \emph{Fluent Expressions} (\codetext{FE}),
which are defined as follows:
\begin{eqnarray}\label{FEdef}
\codetext{FE}
&
::=
&
n
\:\: |\:\: f^t
\:\: |\:\: \codetext{FE}_1 \oplus \,\codetext{FE}_2
\:\: |\:\:
- (\codetext{FE})
\:\: |\:\:
\codetext{abs}(\codetext{FE})
\:\: |\:\:
\codetext{rei}(\codetext{C})
\end{eqnarray}
where $n \in\mathcal{V}$,
$f \in \mathcal{F}$,
$t \in \{0,-1,-2,-3,\dots\}$,
$\oplus \in \{+,-,*,/,\codetext{mod}\}$,
and $r\in \nat$.
$\codetext{FE}$ is referred to as  a \emph{timeless expression} if it
contains no occurrences of $f^{t}$ with $t\neq 0$.
$f$ can be used as a shorthand of  $f^0$.
%
%
The notation $f^t$ is an \emph{annotated} fluent expression.
The expression refers to a relative time reference, indicating  the value
$f$ had $-t$ steps in the past.
The last alternative in (\ref{FEdef}), a \emph{reified expression},
requires the notion of constraint~$\codetext{C}$, introduced below.
$\codetext{rei}(\codetext{C})$ represents a
 Boolean value that reflects the  truth value  of $\codetext{C}$.
%
A \emph{Primitive Constraint} (\codetext{PC})
is  formula
$\codetext{FE}_1 \,\op\, \codetext{FE}_2$,
where $\codetext{FE}_1$ and $\codetext{FE}_2$ are fluent expressions,
and $\op \in \{=, \neq, \geq, \leq, >, <\}$.
A \emph{constraint} $\codetext{C}$ is a
propositional combination of \codetext{PC}s.
We will refer to the primitive constraints of the form
$f = FE$, where $f\in {\cal F}$, as a \emph{basic primitive constraint}.
We accept the constraint
$\codetext{pair}(\codetext{FE}_1,\codetext{FE}_3) =
 \codetext{pair}(\codetext{FE}_2,\codetext{FE}_4)$
  as syntactic sugar of
$\codetext{FE}_1=\codetext{FE}_2$ and
$\codetext{FE}_3=\codetext{FE}_4$.

An axiom of the form
~$\codetext{action }x$
~
in 
$\mathcal{D}_a$, declares that the
action $x \in \mathcal{A}$
is executable by the agent $a$. Observe that the
same action name $x$ can be used for different
actions executable by different agents. This does not cause ambiguity,
since each agent's knowledge is described by its own action theory.
A special action, \codetext{nop},
is  executable by every
agent, and it does not  change any of the fluents.


\begin{example}\label{volley2}
The  actions for each player $A$ of Example~\ref{volley1} are:
\begin{list}{$\bullet$}{\topsep=1pt \parsep=0pt \itemsep=1pt \leftmargin=10pt}
\item $A: \codetext{move}(d)$ one step in direction $d$, where
        $d$ is one of the eight directions: north,
        north-east, east, $\dots$, west, north-west.
\item $A:\codetext{throw}(d,f)$ the ball in direction $d$ (same eight directions as above)
        with strength $f$
        varying from 1 to a maximum throw power ($5$ in our example).
\end{list}
Moreover, the player of each
team is in charge of checking
if a point has been scored (in such case, he whistles).
We write the actions as
\codetext{act([$A$],action\_name)}
and state these axioms:

\codetext{\footnotesize
\renewcommand{\arraystretch}{.8}
\begin{tabular}{lcl}
\multicolumn{1}{l}{action act([A],move(D))}  &:-&
     whiteplayer(A),direction(D).\su\\

\multicolumn{1}{l}{action act([A],throw(D,F))}  &:-&
     whiteplayer(A),direction(D),power(F).\\
\multicolumn{3}{l}{action act([player(white,1)],whistle).\giu}
\end{tabular}
}

\par\noindent where \codetext{whiteplayer},
\codetext{power}, and \codetext{direction} can be
defined as follows:

\codetext{\footnotesize
\begin{tabular}[b]{l}
whiteplayer(player(white,N)) :- agent(player(white,N)). \su\\
power(1). \hspace{.1cm} power(2). \hspace{.1cm} power(3).\hspace{.1cm} power(4).\hspace{.1cm} power(5). \\
direction(D) :- delta(D,\_,\_). ~~
delta(nw,-1,1).  delta(n,0,1).  delta(ne,1,1). \\
delta(w,-1,0).   delta(e,1,0). delta(sw,-1,-1). delta(s,0,-1). delta(se,1,-1). \eoe
\end{tabular}
}
\end{example}

\noindent
The executability of the actions is described by axioms of the form:
\begin{equation}\label{execDef}
\codetext{executable }x
\codetext{ if } \codetext{C}
\end{equation}
where
$x\in {\cal A}$
and \codetext{C} is a
constraint. The axiom states that $x$ is executable only if
 \codetext{C}
is  entailed by the  state of the world.
We  assume that at least one executability
axiom is present for each action;  multiple executability axioms
are treated disjunctively.

\begin{example}\label{volley3}
In our working example, we can state executability as follows:

\codetext{\footnotesize
\begin{tabular}{l}
executable act([player(white,1)],whistle) if [S eq 0] :- build\_sum(S).\su\\

executable act([A],move(D))  if   [hasball(A) eq 0, defense\_time gt 0,\\
\tab  Net lt  x(A)+DX, x(A)+DX leq MX,  1 leq   y(A)+DY, y(A)+DY leq MY]  :-\\
\tab\tab\tab\tab      action(act([A],move(D))), delta(D,DX,DY),\\
\tab\tab \tab\tab         net(Net), linex(MX), liney(MY).\\
executable act([A],throw(D,F))  if  \\
\tab [hasball(A) gt 0,defense\_time eq 0,
      1 leq x(A)+DX*F, x(A)+DX*F leq MX,\\
\tab  1 leq y(A)+DY*F, y(A)+DY*F leq MY]  :-\\
\tab\tab\tab\tab     action(act([A],throw(D,F))), delta(D,DX,DY), linex(MX), liney(MY).\giu~~
\end{tabular}
}

\par\noindent
These axioms state that neither a player nor the ball
can  leave
the field. \codetext{build\_sum} is recursively defined  to return
the  expression:
$\,\codetext{defense\_time} +
  \codetext{hasball}(A_1) + \cdots +
  \codetext{hasball}(A_n) \,$
where $A_1,\dots,A_n$ are  the players (i.e.,  \codetext{player(white,1)} and \codetext{player(black,1)}).
{The operators $=,\neq,\leq,<$, etc. are concretely represented by
\codetext{eq}, \codetext{neq}, \codetext{leq}, \codetext{lt}, respectively.}
\eoe
\end{example}

\noindent The effects of an action execution
are modeled by \emph{dynamic causal laws:}
\begin{equation}\label{dynaLaw}
x \codetext{ causes } \Eff \codetext{ if } \Prec
\end{equation}
where $x \in \mathcal{A}$,  $\Prec$ is a
constraint,
and $\Eff$ is a conjunction of basic primitive constraints.
The axiom asserts that if $\Prec$ is true with
respect to the current state,
then $\Eff$ must hold after the execution of~$x$.

Since agents share  fluents, their actions may interfere
and cause inconsistencies.
A \emph{conflict} happens when the effects of different concurrent actions
are incompatible and would lead to an inconsistent state; note that we allow only consistent
states to exist during the evolution of the world.
A  procedure has to be applied to resolve a conflict and
determine a consistent subset of the conflicting actions
(see Sect.~\ref{sec:inconsistency}).

\begin{example}\label{volley4}
Let us describe the effects
of the actions in the volleyball domain.
When the ball is thrown with force $f$ in direction $d$,
it reaches a destination cell whose distance is as follows:
\emph{a)} if $d$ is either north or south then $\Delta\!X=0, \Delta\!Y=f$;
\emph{b)} if $d$ is east or west then $\Delta\!X=f, \Delta\!Y=0$;
\emph{c)} if $d$ is any other direction, $\Delta\!X=f, \Delta\!Y=f$.
An additional effect is to set  the fluent {\tt defense\_time}
(to $1$ in our example).

\codetext{\footnotesize
\begin{tabular}{l}
act([A],throw(D,F))
causes
hasball(A) eq 0 \hspace{.1cm}
:-
     action(act([A],throw(D,F))).\su\\

act([A],throw(D,F))
causes
defense\_time eq 1
:-
     action(act([A],throw(D,F))).\\

act([A],throw(D,F))
causes
pair(x(ball),y(ball)) eq  pair(x(A)\back + F*DX,y(A)\back + F*DY)
     :-\\
\tab\tab\tab     action(act([A],throw(D,F))), delta(D,DX,DY).\\

     act([A],throw(D,F),
causes
hasball(B) eq 1\\
\tab if       [pair(x(B),y(B)) eq pair(x(A)+F*DX, y(A)+F*DY)] :-\\
\tab\tab\tab     action(act([A],throw(D,F))),
     player(B), neq(A,B),delta(D,DX,DY).\\
act([A],throw(D,F))
causes
point(black) eq 1
if [x(A)+F*DX eq Net]     :- \\
\tab\tab\tab     action(act([A],throw(D,F))), delta(D,DX,\_),
     net(Net).\giu
\end{tabular}
}

\par\noindent
The effects of the other two actions \codetext{move} and
\codetext{whistle} can be stated by:

\codetext{\footnotesize
\begin{tabular}{l}
act([player(white,1)],whistle)
causes
point(white) eq 1
if [x(ball) lt NET] :- \su\\
\tab\tab\tab        net(NET).\\

act([player(white,1)],whistle)
causes
point(black) eq 1
 if [NET lt x(ball)]
:-\\
\tab\tab\tab net(NET).\\

act([A],move(D))
causes
pair(x(A),y(A)) eq pair(x(A)\back+DX,y(A)\back+DY)
      :-\\
\tab\tab\tab      action(act([A],move(D))), delta(D,DX,DY).\\

act([A],move(D)) causes
defense\_time eq defense\_time\back - 1
:-   action(act([A],move(D))).\\

act([A],move(D))
causes
hasball(A) eq 1 \\
\tab
 if  [pair(x(ball),y(ball)) eq pair(x(A)+DX,y(A)+DY)]
      :- \\
\tab\tab\tab      action(act([A],move(D))), delta(D,DX,DY).\eoe
\end{tabular}
}
\end{example}

\noindent
In presence of a conflict (i.e., two agents executing actions that assign a distinct
value to the same fluent),
at least two perspectives
can be followed, by assigning either a passive or an active role to the conflicting agents.
In the first case, a supervising entity is in charge of resolving the conflict,
and all the agents will comply with the supervisor's decisions. Alternatively,
the agents themselves are
in charge of reaching an agreement, possibly through negotiation.
In  the latter case,
the following declarations allow one to specify in the action theories
some basic reaction policies the agents might apply:
\begin{equation} \label{axiom:reaction}
\codetext{action } x \codetext{ [} OPT \codetext{]}
\end{equation}
with $OPT$ defined as:
{$\begin{array}[t]{rcl}
OPT & ::= & \phantom{|} \codetext{ on\_conflict }  OC \;\; \codetext{[} OPT \codetext{]} \\
&& |  \codetext{ on\_failure }  OF  \;\; \codetext{[} OPT \codetext{]}\\
OC & ::=  & \phantom{|}  \codetext{ retry\_after } T \codetext{ [provided } C\codetext{]}\\
&&  | \codetext{ forego [provided } C \codetext{]}\\
OF& ::=  &  \phantom{|}  \codetext{ retry\_after } T \codetext{ [if } C\codetext{]}\\
&&  | \codetext{ replan [if } C \codetext{] [add\_goal } C\codetext{]}\\
&&  | \codetext{ fail [if } C \codetext{]}
\end{array}$}

\noindent where $T$ is a number of steps and $C$ is a constraint.
Notice that one can also specify policies
to be adopted whenever a failure occurs in executing an action.

We remark here the difference between \emph{conflict} and \emph{failure}.
A conflict occurs whenever concurrent actions performed by different agents
try to make inconsistent modifications to the state of the world.
A failure occurs whenever an action~$x$ cannot be executed as planned by an agent~$a$.
This might happen, for instance, because after the detection of a conflict involving~$x$,
the outcome of the conflict resolution phase requires~$x$ to be inhibited.
In this case the agent~$a$ might have to
reconsider its plan. Hence, reacting to a failure is a ``local'' activity
the agent might perform after the state transition has been completed.
In axioms of the form (\ref{axiom:reaction}),
one can specify different reactions to a conflict (resp.~a failure) of the same action.
Alternatives will be considered in their order of appearance.

\begin{example}
Let us assume that the agents $a$ and $b$ have priority $0$, while agent $c$ has  lower priority $2$.
Let us also assume that the current state is such that actions \codetext{act\_a},
\codetext{act\_b}, and \codetext{act\_c} are all executable
(respectively, by agents $a$, $b$, and $c$),
where their effects on fluent $f$ are
of setting it to 1, 2, and 3, respectively. This indicates a situation of
conflict, since the effects of the concurrent execution of the three actions are inconsistent.
Assume that the following options have been defined:

\codetext{\small
\begin{tabular}{l}
\codetext{action  act\_a on\_conflict retry\_after 2}\\
\codetext{action  act\_b on\_conflict forego}\\
\codetext{action  act\_c on\_failure retry\_after 3}\\
\end{tabular}
}

\noindent
and that the plan of agent $a$ (resp., $b$, $c$) requires the execution of action \codetext{act\_a}
(resp.,  \codetext{act\_b},  \codetext{act\_c}) in the current state.
One possible conflict resolution
 is to focus  the  priority of the  agents.
This causes  \codetext{act\_c} to be removed from the execution list.
Thus, agent $c$ fails in executing \codetext{act\_c} and will retry the same action after 3 steps.

Some policy must be now chosen to resolve the conflict between $a$ and $b$.
The first possibility is that agents have passive roles in conflict resolution,
and a supervisor selects, according to some criteria,
a  consistent subset of the actions/agents.  For example, if
$a$ is selected (e.g., by  lexicographic order), then the state
will be modified by setting $f=1$, declaring  \codetext{act\_a} successful, while
agent $b$ will fail.

An alternative  is to  allow
the agents $a$ and $b$ to directly resolve the conflict, using  their \codetext{on\_conflict} options. This causes $a$
to retry the execution of  \codetext{act\_a} after 2 time steps and $b$ to forego the
execution of  \codetext{act\_b}. Both of them
will get a failure message, because neither \codetext{act\_a} nor \codetext{act\_b}
are executed.
\eoe\end{example}
Apart from the possible communications occurring among agents during the
conflict resolution phase,
other forms of ``planned'' communication
can be modeled in an action theory.
An axiom of this form

\begin{equation}\label{ax:broadcast}
\codetext{request } C_1 \codetext{ if } C_2
\end{equation}
describes a special static causal law that allows an agent to broadcast
a request, whenever a certain condition ($C_2$) is
encountered. By executing this action,
an agent asks if there is another agent that can make
the constraint $C_1$ true. Only an agent knowing all of the fluents
occurring in $C_1$ is allowed to answer such request.


Instead of broadcasting an help request, an agent $a$ can send such a message
directly to another agent by providing its name:\footnote{Any
request sent to a nonexistent agent will never receive an answer.}
\begin{equation}\label{ax:p2p}
\codetext{request } C_1 \codetext{ to\_agent } a' \codetext{ if } C_2
\end{equation}

The following  communication primitive
 subsumes the previous ones:
\begin{eqnarray}\label{ax:comm}
\codetext{request } C_1 \codetext{[ to\_agent } a' \codetext{] if }
C_2 \codetext{  [ offering } C_3 \codetext{ ]}
\end{eqnarray}
If the last option is used, the requesting agent also provides a
``reward'' by promising to ensure
$C_3$ in case of acceptance of the proposal.
Axioms of this type allow us to model negotiations and other forms of
 bargaining and transactions.

 In turn, agents may declare their willingness to accept requests and serve
 other agents using statements of the form
 \begin{eqnarray}\label{ax:comm_ans}
\codetext{help } Agent\_List \:\codetext{[ if } C \codetext{]}
\end{eqnarray}
where $Agent\_List$ is either a list of agent names $a_1, \dots, a_k$---denoting that the agent in question can serve requests coming
from the agents $a_1,\dots, a_k$---or the
keyword $all$---denoting the fact that the agent can accept
requests coming from any source.  The optional condition allows the
agent to select which requests to consider depending on properties
of the current state of the world.

\begin{example}
Let us consider a domain with  three agents: a guitar maker,
a joiner that provides wooden parts of guitars (bodies and necks),
and a seller that sells strings and pickups.
We assume that the maker has plenty of money (so we do not take
into account what it spends), that
the seller wants to be paid for its materials,
and that necks and bodies can be obtained for free
(e.g., the joiner has a fixed salary paid by the maker).
The income of the seller is modeled by changes to the
value of the fluent \codetext{seller\_account}.
In Figure~\ref{Fig:liutaio} we report
an action description that models the agent
\codetext{guitar\_maker}---analogous theories can be formulated for the other two agents.
Observe that two point-to-point
interactions are modeled---namely,
the one between the \codetext{guitar\_maker} and the \codetext{joiner}, to obtain necks and bodies,
and the one between
the \codetext{guitar\_maker} and the \codetext{seller}, to buy strings (\$8) and pickups (\$60).
Two kind of guitars can be made, differing in the number of pickups.
\eoe

\begin{figure}[tb]
\centerline{\fbox{\begin{minipage}{0.95\textwidth}
{\centerline{\scriptsize$\begin{array}{l}
\codetext{agent  guitar\_maker.}\\
\codetext{action make\_guitar.}
\\
\codetext{executable make\_guitar if  neck} > 0 \codetext{ and strings} >= 6 \codetext{ and }
                   \codetext{body} > 0 \codetext{ and pickup} > 0\codetext{.}
\\[\spv]
\codetext{\% actions for making two different kinds of guitars:}\\
\codetext{make\_guitar causes~ guitars=guitars$^{-1}$+1  ~and~ neck=neck$^{-1}$-1 ~and~  body=body$^{-1}$-1}
\\\phantom{\codetext{make\_guitar causes~ }}
     \codetext{and~ strings}=\codetext{strings}^{-1}-6 \codetext{ ~and~ }
     \codetext{pickup}=\codetext{pickup}^{-1}-2
\\\phantom{\codetext{make\_guitar causes~ }}
   \codetext{if pickup}>=2\codetext{.}
\\
\codetext{make\_guitar causes guitars=guitars$^{-1}$+1  ~and~ neck=neck$^{-1}$-1 ~and~}
      \codetext{strings} = \codetext{strings}^{-1}-6
\\\phantom{\codetext{make\_guitar causes~ }}
      \codetext{and~ body=body$^{-1}$-1  ~and~ pickup=pickup$^{-1}$-1}\\
\phantom{\codetext{make\_guitar causes~ }}
  \codetext{if pickup} < 2\codetext{.}
\\[\spv]
\codetext{\% interaction with joiner:}\\
\codetext{request neck}>0  \codetext{ to\_agent joiner if neck} = 0\codetext{.}
\\
\codetext{request body}>0  \codetext{ to\_agent joiner if body} = 0\codetext{.}
\\[\spv]
\codetext{\% interaction with seller:}\\
\codetext{request strings}>5 \codetext{ to\_agent seller if strings}<6
\\\phantom{\codetext{request strings}>5 ~}
      \codetext{offering seller\_account}=\codetext{seller\_account}^{-1}+8\codetext{.}
\\
\codetext{request pickup}>0 \codetext{ to\_agent seller if pickup}=0
\\\phantom{\codetext{request pickup}>0 ~}
      \codetext{offering seller\_account}=\codetext{seller\_account}^{-1}+60\codetext{.}
\\[\spv]
\codetext{\% the goal is to make 10 guitars:}\\
\codetext{goal guitars}=10\codetext{.}
\\[\spv]
\codetext{\% initially the maker owns some material:}\\
\codetext{initially }
  \codetext{guitars} = 2 \codetext{ and } \codetext{body} = 3    \codetext{ and }
  \codetext{neck} = 5    \codetext{ and }
  \codetext{pickup} = 6  \codetext{ and } \codetext{strings} = 24\codetext{.}
\end{array}
$}}\end{minipage}}}
\caption{\label{Fig:liutaio}An action description in \name\ for a guitar maker agent}
\end{figure}
\end{example}

Various forms of
\emph{global constraint} can be exploited to impose control knowledge and
maintenance goals.
These constraints represent properties that must
always persist in the world where the agents act.
Some examples:
\begin{itemize}
\item
$FC\;~\codetext{holds\_at}~\;n$. This constraint
is satisfied if the
fluent constraint $FC$ holds at the $n^{th}$ time step.
\item
$\codetext{always}\:~FC$. This constraint
imposes the condition that the fluent constraint $FC$
holds in all the states of the evolution of the world.
\end{itemize}
Semantics of these constraints is reported in Section~\ref{sub:semantics}.

An \emph{action domain description} consists of  a collection
$\mathcal{D}_a$ of axioms of the forms described so far, for each agent
$a \in \mathcal{G}$.
Moreover it includes, for each agent $a$,  a collection
$\mathcal{O}_a$ of goal axioms (objectives), of the form
$\codetext{goal } \codetext{C}
$,
where $\codetext{C}$ is a constraint,
and a
collection $\mathcal{I}_a$ of initial state axioms
of the form:
$
\codetext{initially } \codetext{C}
$, where $\codetext{C}$ is a constraint involving
only timeless expressions.
For the sake of simplicity, we assume that  all
 the sets $\mathcal{I}_a$ are drawn from a consistent
global initial state description $\mathcal{I}$, i.e.,
$\mathcal{I}_a \subseteq \mathcal{I}$.
A specific instance of a planning problem
is a triple
$$
\left\langle
 \left\langle \mathcal{D}_a \right\rangle_{a\in\mathcal{G}},
 \left\langle \mathcal{I}_a \right\rangle_{a \in \mathcal{G}},
 \left\langle\mathcal{O}_a \right\rangle_{a \in \mathcal{G}}
\right\rangle{\texttt{.}}
$$
%


\section{System behavior}\label{sec:semantics}

The behavior of \name\ can be split into two parts:
the semantics of the action description language, parametric on the
supervisor selection strategy, and these strategies that can be programmed.
We present the former in Section~\ref{sub:semantics}, the latter in
Sections~\ref{planExec}--\ref{sec:message}.
Finally, some implementation notes are reported in Section~\ref{sec:implementation}.

\subsection{Semantics of \name}\label{sub:semantics}

The semantics of the  action language is described by a transition function that
operates on states.
A state $s$ is identified by a total function
$v : \mathcal{F}  \longrightarrow \mathcal{V}$.
We assume a given  horizon  $\N$,
within which the planning activities of all
agents have to be completed.

Let $\vec{v} = \langle v_0,\dots,v_i \rangle$ be a  state sequence, with $0 \leq i \leq \n$.
Given $\vec{v}$,  $j \in \{0,\dots, i\}$, and a fluent expression $\varphi$,
we define the concept of \emph{value} of $\varphi$ in  $\vec{v}$ at time $j$,
denoted by $\vec{v}(j,\varphi)$, as follows:
$$
\begin{array}{rcl}
\vec{v}(j,x) &=& x  ~~ \mbox{ if $x \in \mathcal{V}$}
\\
\vec{v}(j,f^t) &=& v_{j+t}(f) ~~ \mbox{ if $f \in \mathcal{F}$, and $0\leqslant j+t$}
\\
\vec{v}(j,f^t) &=& v_0(f) ~~ \mbox{ if $f \in \mathcal{F}$, and $j+t < 0$}
\\
\vec{v}(j,\codetext{abs}(\varphi)) &=& |\vec{v}(j,\varphi)|
\\
\vec{v}(j,-(\varphi)) &=& -(\vec{v}(j,\varphi))
\\
\vec{v}(j,\varphi_1 \oplus \varphi_2)&=& \vec{v}(j,\varphi_1) \oplus \vec{v}(j,\varphi_2)
\\
\vec{v}(j, \codetext{rei}(C))&=& 1 ~~  \mbox{if $\vec{v}\models_j C$}
\\
\vec{v}(j, \codetext{rei}(C))&=& 0 ~~  \mbox{if $\vec{v}\not\models_j C$}
\end{array}
$$
where $\oplus \in \{+,-,*,/,\codetext{mod}\}$. The last two cases specify the semantics
of reification that relies on the
notion of \emph{satisfaction}, which in turn is defined by structural induction on
constrains, as follows.
Given a primitive constraint $\varphi_1 \:\codetext{op}\:\varphi_2$ and a state sequence
$\vec{v}$,
the notion of satisfaction at time $j$ is defined as:
$  \vec{v} \models_j  \varphi_1  \:\codetext{op}\:\varphi_2$ iff $\vec{v}(j,\varphi_1) \:\codetext{op}\:  \vec{v}(j,\varphi_2)
$.
The notion $\models_j $ is generalized to the case of
propositional combinations of  fluent constraints in the usual manner.
For the case of $\codetext{pair}$, we have  that
$\vec{v}\models_j \codetext{pair}(E_1,E_3)=  \codetext{pair}(E_2,E_4)$
if and only if
$\vec{v}\models_j E_1=E_2 \wedge E_3=E_4$.

\smallskip

We recall that a \emph{timeless fluent} is a fluent expression of the form $f^0$ (and $f$).

Given a constraint $C$ and a state sequence $\vec{v} = \langle v_0,\dots,v_i\rangle$,
let $\FV(C)$ be the set of timeless fluents occurring in $C$.
A function $\sigma:\FV(C) \longrightarrow \mathcal{V}$
is a $\vec{v}$-\emph{solution} of $C$ if $\langle v_0,\dots,v_i,\sigma\rangle \models_{i+1} C$.
Let us observe that this definition makes use of a slight abuse of notation, since $\sigma$ is
potentially not a complete state (some fluents may have not been assigned a value by $\sigma$).
Nevertheless, the choice of fluents in $\FV(C)$ guarantees the possibility of correctly
evaluating $C$. In other words, $\sigma$ can be seen as a partial state contributing (with
$\vec{v}$) to the satisfaction of  $C$ at time $i+1$.
Let us see how to complete this state using inertia:
if $\sigma$ is a $\vec{v}$-solution of a constraint $C$,
$\ine(\sigma,\vec v)$ is defined as follows:
$$\ine(\sigma,\vec{v})(f) = \left\{ \begin{array}{ll}
          \sigma(f) & \mbox{if $f \in \FV(C)$}\\
          {v}_i(f) & \mbox{otherwise}
          \end{array}
          \right.$$
Fluents not appearing in $C$ are considered inertial (namely they
maintain their previous values) and therefore the state is completed
using the function $\ine$.

An action $x$ is \emph{executable} by agent $a$ in a state sequence
$\vec{v} = \langle v_0,\dots,v_i\rangle$
if there is at least an  axiom
~$\codetext{executable }x\codetext{ if }C$~ in $\mathcal{D}_a$
and it holds that $ \vec{v}\models_i  C$.
If there is more than one executability condition,
it is sufficient for  one of them to apply.

Let us denote with $Dyn(x)$ the set of dynamic causal law axioms for
action $x$.
The \emph{desired effect} of executing $x$ in state sequence $\vec{v} = \langle v_0,\dots,v_i\rangle$,
denoted by $\DEff(x, \vec{v})$, is a constraint defined as follows:
\[ \DEff(x, \vec{v}) =\bigwedge \left\{ \Eff \:|\:
        x \codetext{ causes } \Eff \codetext{ if }\mathit{Prec} \in Dyn(x),
         \vec{v}\models_i  \mathit{Prec}\right\}. \]
Request accomplishment actions can be used in the construction of this set.

Given a state sequence $\vec{v} = \langle v_0,\dots,v_i\rangle$, a state $v_{i+1}$,
and a set of actions $X$,
a triple $\langle \vec{v},X,v_{i+1}\rangle$
is a \emph{valid state transition} if:
\begin{itemize}
\item for all $x \in X$, the action $x$ is executable in $\vec{v}$ by some agent $a$, ~and

\item
$v_{i+1} = \ine(\sigma,\vec{v})$,
where $\sigma$ is a $\vec{v}$-solution of the constraint
${\bigwedge_{x \in X}}\DEff(x,\vec{v})$.
\end{itemize}
Observe that if $X=\emptyset$, then $\langle \vec{v},\emptyset,v_i\rangle$
will be a valid state transition.

Let  $\vec{v} = \langle v_0, \dots, v_{\n}\rangle$ be a sequence
of  states,
$\langle (\mathcal{D}_a)_{a \in \mathcal{A}}, (\mathcal{I}_a)_{a \in \mathcal{A}},
(\mathcal{O}_a)_{a \in \mathcal{A}}\rangle$
 an instance of a planning problem,
 and $X_1,\ldots,X_{\n}$ be sets of actions.
We say that $\langle v_0,X_1,v_1, \dots, X_{\n},v_{\n}\rangle$
is a \emph{valid trajectory} if:
\begin{itemize}

\item for each agent $a$ and for each axiom of the form $\codetext{initially } C $ in $\mathcal{I}_a$,
we have that $\vec{v} \models_0~C$,
\item for all $i \in \{0,\dots,\n-1\}$,
$\langle \langle v_0,\dots,v_i\rangle,X_{i+1},v_{i+1}\rangle$ is a valid state transition.
\end{itemize}

A valid trajectory is \emph{successful} for an agent $a$ if, for each axiom of the form $\codetext{goal }  C $ in $\mathcal{O}_a$,
it holds that $\vec{v} \models_{\n} C$.

At each time step $i$, each agent might propose a set of actions
for execution---we assume that all the  proposed actions are  executable in the
state sequence $\vec{v}_i = \langle v_0,\dots,v_i\rangle$. Let $Y_{i+1}$ be this set of  actions.
The supervisor selects a subset $X_{i+1} \subseteq Y_{i+1}$ such that
the constraint $\Eff(X_{i+1},\vec{v}_i)$, defined as:
$$\Eff(X_{i+1},\vec{v}_i) =\bigwedge_{x \in X_{i+1}} \DEff(x,\vec{v}_i)$$
is satisfiable w.r.t. $\vec{v}$---i.e., there exists a complete state $v_{i+1}$ such
that $\langle \vec{v}_i, X_{i+1}, v_{i+1}\rangle$ is a valid state transition.
It is the job of the supervisor to determine the subset $X_{i+1}$ given
$Y_{i+1}$ and $\vec{v}_i$---as a maximal consistent subset, using
agent priorities or other approaches, as discussed in
Section~\ref{sec:inconsistency}. If an agent cannot find a plan at the time step $i$
it will ask for a \codetext{nop} and try again the next step.

\medskip

Let us complete the semantics of the language by dealing with request and help laws.
A request of the agent $a$
$$\codetext{request } C_1 \codetext{ to\_agent } a' \codetext{ if } C_2$$
is \emph{executable} in a state sequence $\vec{v} = \langle v_0,\dots,v_i\rangle$
if it holds that $\vec{v}\models_i   C_2 $. If the request above is executable,
it can be \emph{accomplished} in the successive state $v_{i+1}$ if
there is an axiom
$$\codetext{help }  \cdots a \cdots \codetext{ if } C_3$$
in $\mathcal{D}_{a'}$
and $ \langle v_0,\dots,v_i,v_{i+1}\rangle \models_{i+1}  C_3$.
The semantics of the help law is that of enabling a  request accomplishment (after a request demand)
and it can be viewed as the execution of an ordinary action
by agent $a'$.\footnote{We hypothetically assume that $a'$ has access
to all fluents of $a$.}
We can view this as if $a'$ had an additional
action  $y$ defined in $\mathcal{D}_{a'}$ as:
$$\begin{array}{c}
\codetext{executable } y \codetext{ if } C_3 \wedge (C_2)^{-1}
\\
y\codetext{ causes } C_1 \codetext{ if } \codetext{true}
\end{array}$$
Observe that, as happens for executability laws, multiple help preconditions are
considered disjunctively.
If the request includes also the option \codetext{offering $C_4$}, then
the action $y$ will cause $C_1 \wedge C_4$ as effect.

Let us add some comments on agents' requests for action execution.
Each agent wishes to execute some actions and
to ask some requests. After the supervisor has decided which actions
will be executed, each agent retrieves the relevant requests and
analyzes them in order to possibly fulfill them in the next time step
(see below for further details).
These requests behave like an action $y$, as stated above.


Two global constraints are allowed by the language \name. Their effect is to
filter out sequences of states that do not fulfill those constraints:
\begin{itemize}
\item $C \codetext{ holds\_at } i$ imposes that any valid trajectory
$\langle v_0,X_1,v_1, \dots, X_{\n},v_{\n}\rangle$
must satisfy $\langle v_0,v_1, \dots, v_{\n}\rangle \models_i C$

\item $\codetext{always } C$ imposes that any valid trajectory
$\langle v_0,X_1,v_1, \dots, X_{\n},v_{\n}\rangle$
must satisfy $\langle v_0,v_1, \dots, v_{\n}\rangle \models_i C$
for all $i \in \{0,\dots,\n\}$.

\end{itemize}
The supervisor is in charge of checking if these constraints can be
satisfied while selecting $X_i$ as mentioned before.
If the fluents involved in the constraints are all
known to an agent $a$, the set of actions proposed by $a$
are such that they will guarantee the property if all of them
(and only them) are selected for application.


Each  agent $a$, at each time step $i$,
selects a set of actions $Y^a_{i+1}$ it wishes to execute.
For doing that, $a$ looks for a sequence of (sets of)
actions to achieve its local goal, given the
current state sequence $\langle v_0,\dots,v_i\rangle$.
The set of actions $Y^a_{i+1}$ are those to be executed at the current time step.
If the new state $v_{i+1}$ communicated by the supervisor
is different from the
state it expected after the application of all the actions in
the set $Y^a_{i+1}$ (due either to the fact that some of these actions
are not selected, or that other agents have executed actions that
have unexpectedly changed some values), it will need to replan.
Let us observe that, although globally the supervisor views a valid
trajectory, locally this is not true (some state transitions are
not justified by the actions of agent $a$ alone). However,
in looking for a plan (and in replanning), it reasons on an
``internal'' valid trajectory from the current time to the future.

\smallskip

Let us focus on the problem of reacting to \codetext{request}s.
Suppose that an agent $a'$, at time $i$  in a state sequence
$\langle v_0,v_1, \dots, v_{i}\rangle$, receives the requests
$r_1,\dots,r_h$, where $r_j$ is of the form
$$\codetext{request }C^j_1\codetext{ to\_agent }a'\codetext{ if }C^j_2$$
and, moreover, assume that these requests are ordered
(e.g., by the priorities of the requesting agent $a_j$).
For $j=1,\dots,h$, if $\mathcal{D}_{a'}$ contains an axiom
$$\codetext{help}\cdots a_j\cdots\codetext{ if }C^j_3$$
such that $\langle v_0,v_1, \dots, v_{i}\rangle \models_i C^j_3$,
the agent $a'$ adds temporarily to its theory the constraint
\begin{equation}\label{conagg}
C^j_1\codetext{ holds\_at }i+1
\end{equation}
and looks for a plan in the enlarged theory.
If such a plan exists, the constraint (\ref{conagg}) is definitely stored
in $\mathcal{D}_{a'}$, otherwise the request is ignored.
In both cases, $a'$
proceeds with next request ($j:=j+1$).
At the end, some (possibly none) of the $h$ constraints
$C^j_1,\dots,C^j_h$
will
be fulfilled by a plan and the set of actions $Y^{i+1}_{a'}$
of the next step of this plan are passed to the supervisor.

\smallskip

Let us focus now on how the agent $a$ deals with the options related to a failure
(this is also developed in Section~\ref{sub:failure}).
Let us assume an action $x$ submitted for execution at time $i$ has not been
 selected by the supervisor, and, therefore, a failure signal is returned
to the agent $a$. The current sequence of states is
$\vec v=\langle v_0,v_1, \dots, v_{i+1}\rangle$.

Let us analyze what happens in the three options:
\begin{itemize}
\item  $\codetext{fail  if } C_1 $:
if $\vec{v} \models_{i+1} C_1$ then agent $a$ declares its failure. From this point onwards,
the agent will not generate any actions, nor interact with other agents.

\item $\codetext{replan if } C_1 \codetext{ add\_goal } C_2$:
if $\vec{v} \models_{i+1} C_1$ then $\codetext{goal } C_2$ is added in $\mathcal{D}_a$
(and then the agent $a$ starts replanning)

\item $\codetext{retry\_after } T \codetext{ if } C_1$:
if $\vec{v} \models_{i+1} C_1$ then
for $T-1$ time steps the agent $a$ requires only \codetext{nop} to the supervisor,
at time step $T+i$ the action $x$ is required again.
\end{itemize}

If the \codetext{if} option is missing, the condition will be assumed to be satisfied.
If the \codetext{add\_goal} option is missing, no new goal will be added.


\subsection{Concurrent plan execution}\label{planExec}

The agents are autonomous and develop their activities independently,
 except for the execution of the actions/plans.
In executing their plans, the agents must take into account the effects of
concurrent actions.

We developed the basic communication mechanism among agents
 by exploiting
a \emph{tuple space}, whose access and manipulation follows the
blackboard principles introduced in the Linda model \cite{linda}.
Linda
is a popular model for coordination and communication among processes; Linda offers
coordination via a shared memory, commonly referred to as a
\emph{blackboard} or
\emph{tuple-space}. All the information are stored in the blackboard in the form of
\emph{tuples}---the shared blackboard provides atomic access and associative memory behavior (in retrieving
and removing tuples). The SICStus Prolog implementation
of Linda allows the definition of a server process, in charge of managing the
blackboard, and client processes, that can add tuples (using the {\tt out} operation),
read tuples (using the {\tt rd} operation) and remove tuples (using the {\tt in} operation).

Most of the interactions among concurrent agents, especially
those interactions aimed at resolving conflicts, are managed by
a specific process, the \emph{supervisor}, that also provides a global
time to all agents, enabling them to execute their actions synchronously.
The supervisor process stores the initial state
and the changes caused by the successful executions of actions.
It synchronizes the actions execution, and controls the coordination and the
arbitration in case of conflicts. It also sends a success or a failure signal
to each agent at each action execution attempt, together with the list of
changes to its local state.

Let us describe how the execution of concurrent plans proceeds.
As mentioned, each action description includes a set of constraints
describing a portion of the initial state.

\begin{enumerate}
\item At the beginning,
the supervisor acquires the specification
$\mathcal{I}=\bigcup_{a\in\mathcal{G}}\mathcal{I}_a$
of the initial state.

\item At each time step the supervisor starts a new state transition:
\begin{list}{$\bullet$}{\topsep=1pt \parsep=0pt \itemsep=1pt \leftmargin=12pt}
\item Each
 agent sends to the supervisor a request to perform an action---i.e.,
the next action of its locally computed plan---by specifying
its effects on the (local) state.

\item The supervisor collects all these requests and starts an analysis, aimed at
determining the subsets of actions/agents that conflict (if any).
A conflict occurs whenever agents require
incompatible assignments of values to the same fluents.
The transition takes place once all conflicts have been resolved and a subset
of compatible  actions has been identified by means of some
policy (see below). These actions are enabled while the remaining ones
are inhibited.

\item All the enabled actions are executed, producing
 changes to the global state.

\item These changes are then sent back to all agents, to achieve the
	corresponding updates of each agent's local state.
All agents are also notified about the outcome of the procedure.
In particular, those agents whose actions have been inhibited receive a failure
message.
\end{list}

\item The computation stops when the time $\N$ is reached.
\end{enumerate}
Observe that, after each step of the local plan execution,
each agent needs to check if the reached state still supports its successive
 planned
actions. If not, the agent has
 to reason locally and revise its plan, i.e., initiate a replanning phase.
 This is due
to the fact that the reached state might be different from
the expected one. This may occur in two cases:
\begin{enumerate}
\item
The proposed action was inhibited, so the agent actually executed a \codetext{nop}; this
case occurs when the agent receives a failure message from the supervisor.

\item
The interaction was successful, i.e., the planned action was executed,
but the effects of the actions performed by other agents
affected fluents in its local state, preventing the successful continuation
of the remaining part of the local plan. For
 instance, the agent $a$ may have assumed that the fluent $g$
maintained its value by inertia, but another agent, say $b$, changed
such value. There is no direct conflict between the actions of $a$ and $b$,
but agent $a$ has to verify that the rest of its plan is still applicable (e.g.,
the next action in $a$'s plan may have lost its executability condition).
\end{enumerate}

\subsection{Conflict resolution}\label{sec:inconsistency}

A conflict resolution procedure is invoked by the supervisor whenever
it determines the presence of a set of incompatible actions.
Different policies can be adopted in this phase and different roles can be played
by the supervisor.

First of all, the supervisor exploits the  priorities of
the agents to attempt a resolution of the conflict,
 by inhibiting the actions issued by  low priority agents.
If this does not suffice, further options are applied.
We describe here some of the easiest viable possibilities,
that we have already implemented in our prototype.
The architecture of the system is  modular (see
 Sect.~\ref{sec:implementation}),
and  can be easily extended
to include more complex policies and protocols.

The two approaches we implemented so far differ by assigning the active role in
resolving the conflict
either \emph{(a)} to the supervisor or \emph{(b)} to the conflicting agents.

\smallskip

\noindent
In the {\em first case},
the supervisor has an \emph{active role}---it acts as a referee and decides,
without any further interaction with the agents,
which actions have to be inhibited.
In the current prototype, the arbitration strategy is limited  to:
\begin{itemize}
\item
A random selection of a single action to be executed; or
\item
The computation of a maximal set of compatible actions to be executed. This
computation is done by solving a CSP---which is dynamically generated using
a   CLP($\mathcal{FD}$) encoding.
\end{itemize}
Note that, in this strategy, the
\codetext{on\_conflict} policies assigned to actions by axioms~(\ref{axiom:reaction})
are ignored.
This  ``centralized'' approach is relatively simple; it  has also strong potential of
facilitating the creation of optimal plans. On the other hand,
the adoption of a centralized approach to conflict resolution might become
a bottleneck in the system, since
all conflicting agents must wait for supervisor's decisions.

\smallskip

\noindent
In the {\em second case},
the supervisor simply  notifies the set of conflicting agents
about the inconsistency of their actions.
The  agents involved in the conflict are completely in charge of resolving it
by means of a negotiation phase. The supervisor
waits for a solution from the agents.
In solving the conflict, each agent $a$ makes use of
one of the \codetext{on\_conflict} directives
(\ref{axiom:reaction})
specified for its conflicting action $x$.
The semantics of these directives are as follows
(in all the cases \codetext{[provided $C$]} is an optional qualifier;
if it is omitted it is interpreted as \codetext{provided true}):
\begin{list}{$\bullet$}{\topsep=1pt \parsep=0pt \itemsep=1pt}
\item
The option \codetext{on\_conflict forego provided $C$} causes the agent $a$ to
``search'' among the other conflicting agents for someone, say $b$, that can guarantee the
condition \codetext{$C$}.
In this case,
$b$ performs its action while
the execution of $a$'s action fails, and  $a$
executes a \codetext{nop} in place of its action $x$.
Different strategies can be implemented in order to  perform such a ``search for help''.
A simple one is the  round-robin policy described below, but
many other alternatives are possible and should be considered in completing
the prototype.

\item
The option
 \codetext{on\_conflict retry\_after $T$ provided $C$},
will cause $a$ to execute  \codetext{nop}
during the following $T$ time steps and then it will try again to execute
its action (if the preconditions still hold).
\item
If there is no applicable option (e.g., no option is defined
or none of the agents accept to guarantee \codetext{$C$}),
the action is inhibited and its execution fails.
\end{list}
The way in which agents negotiate and exploit the \codetext{on\_conflict} options
can rely on several  protocols, of different complexity.
For instance, one possibility might be to  nominate  a ``leader'' within each of
the conflicting sets $S$ of agents. The leader is  in charge of coordinating the agents in
$S$  to resolve the conflict without interacting with the supervisor.

Another approach consists of
letting each agent in $S$ free to proceed and to find an agreement by sending proposals
to other agents (possibly by adopting some order of execution, some priorities, etc.)
and receiving their proposals/answers.
In the current prototype, we implemented a round-robin policy.
Let us assume that the state sequence already constructed is $\vec{v}=\langle v_0,\dots, v_i\rangle$
and let us assume that the  agents in the list $A = \langle a_1, \dots, a_m\rangle$ aim at executing
the set of actions $Y=\langle y_1,\dots,y_m\rangle$, respectively.
Furthermore, let us assume that the execution of all actions in $Y$ will introduce a constraint that does not have a
$\vec{v}$-solution.
There is a sorting of the agents, and they take turn in
resolving the conflict.
Suppose that at a certain round of the procedure the agent $a_k$ is selected.
$a_k$ tries its next unexplored  \codetext{on\_conflict OP provided $C$} option
for its action and checks if $\vec{v} \models_i C$.
\begin{itemize}
\item If $\vec{v} \models_i C$ then $a_k$ will apply the \codetext{OP} option and
$a_k$ and $y_k$ are removed from $A$ and $Y$, respectively.

\item Otherwise, the next agent is selected and the successive call to $a_k$ will consider the next
\codetext{on\_conflict} option.
\end{itemize}
If there are no successive options for $a_k$ then $a_k,y_k$ will be removed from $A, Y$
and a failure for $a_k$ will occur.
After each step, if $Y$ has a   $\vec{v}$-solution, then the procedure will terminate and the actions in $Y$
will be executed.
Observe that this procedure always terminates with a solution to the conflict, since
a finite number of \codetext{on\_conflict} options are defined for each action.

This  a relatively rigid policy, and it represents a simple example of how to realize a
 terminating
protocol for conflict resolution. Alternative solutions can be added to the prototype
thanks to its modularity.

\smallskip

Once all conflicts have been addressed, the supervisor applies the
enabled actions, and obtains the new global state. Each agent
receives a communication containing
the outcome of its  action execution and
the changes to its local state.
Moreover, further information might be sent to the participating agents,
depending on the outcome of the coordination procedure. For instance,
when two agents agree on an \codetext{on\_conflict} option, they ``promise''
to execute specific  actions
(e.g., the fact that one agent has to execute $T$ consecutive \codetext{nop}).

\subsection{Failure policies}\label{sub:failure}
Agents receive a failure message from the supervisor whenever their requested
actions have been  inhibited.
In such a case, the original plan of the agent has to be revised to detect if
the  local goal can still be reached, possibly by replanning.
Also in this case different approaches can be applied.
For instance, one agent could avoid developing  an entire
plan at each step, but limit itself to produce a partial plan for the very next step.
Alternatively,  an agent could attempt to determine the
``minimal'' modifications to the existing plan in order to make it
valid with respect to the new encountered state.\footnote{At this time,
the prototype includes only replanning from scratch at each step.}

In this replanning phase, the agent can exploit the \codetext{on\_failure}
options associated to the corresponding  inhibited action.
The intuitive semantics of these options can be described as follows.
\begin{itemize}
\item  \codetext{retry\_after $T$ [if $C$]}:
the agent first evaluates the constraint $C$; if $C$ holds,
then it executes the action \codetext{nop} $T$ times  and then
tries again the failed action (provided that its executability conditions
still hold).

\item  \codetext{replan [if $C_1$] [add\_goal $C_2$]}:
the agent first evaluates $C_1$; if it holds, then in the
following replanning phase the goal $C_2$  is added to the
current local goal.
The option \codetext{add\_goal $C_2$} is optional; if it is not present
then  nothing is added to the goal, i.e., it is the same as
\codetext{add\_goal true}.

\item   \codetext{fail [if $C_1$]}: this is analogous to
\codetext{replan [if $C_1$]} \codetext{add\_goal false}.
In this case the agent declares that it is impossible
to reach its goal. It quits and does not participate to the subsequent
steps of the concurrent plan execution.

\item   If none of the above  options
is applicable,
then the agent will proceed as if the  option \codetext{replan if true} is
present.
\end{itemize}
All the options declared for the inhibited action are considered in the given order,
executing the first applicable one.

It might be the case that some global constraints
(such as $\codetext{holds\_at}$ and
$\codetext{always}$, cf., Sect.~\ref{sec:suntax})
 involve fluents
that are not  known by any of the agents.
Therefore, none of the agents can consider such constraints
while planning.
Consequently, these constraints have to be enforced while merging
the individual plans. In doing this, the supervisor adopts the
same strategies
introduced to deal with conflicts and failures among actions, as
described earlier.
Namely, whenever a global constraint would be violated by the concurrent execution
of actions (taken from different agents' plans) a conflict is generated and
a conflict resolution procedure executed. Thus,
some of the conflicting actions will be inhibited causing their failure.

\subsection{Broadcasting and direct requests}\label{sec:message}


Let us describe a simple protocol for
implementing the point-to-point and broadcast
communications among agents,
following an explicit request of the form~(\ref{ax:comm}).
In particular, let us
assume that the current state is the $i$-th one of the plan execution---hence,
 the supervisor is  coordinating the transition to the
$(i{+}1)$-th state by executing the $(i{+}1)$-th action of each local plan.
The handling of requests is interleaved with the agent-supervisor interactions
that realize plan execution; nevertheless, the supervisor does not intervene on the
requests, and
the requests and offers are directly exchanged among agents.
We can sketch the main steps involved in a state transition, from
the point of view of an agent $a$, as follows:
\begin{enumerate}
\item[(1)]  Agent $a$ tries to execute its action and sends this information
to the supervisor
(Sect.~\ref{planExec}).

\item[(2)]
Possibly after a coordination phase, $a$ receives from the supervisor the outcome
of its attempt to execute the action
(failure or success, the changes in the state, etc.)

\item[(3)]
If the action execution is successful,
before declaring the current transition completed, the agent
$a$ starts an interaction with the
other agents to handle pending requests.
All the communications associated
to such interactions are realized using Linda's
tuple-space (requests and offers are posted and retrieved by
agents).
\begin{list}{$\bullet$}{\topsep=1pt \parsep=1pt \itemsep=2pt}
\item[(3.a)]
Agent $a$ fetches the collection $H$ of all the requests still pending
 and generated until step~$i$.
For each request of help $h\in H$, originating from some agent $b$,
agent $a$ decides whether to accept $h$ or not. Such a decision might involve
 planning activities, in order to determine if the
requested condition can be achieved by $a$, possibly by modifying its
original  plan.
In the positive case, $a$ posts its offer into the tuple-space and waits for a
rendez-vous with~$b$.
\item[(3.b)]
Agent $a$ checks whether there are  replies to the requests it  previously posted.
For each request for which replies are available, $a$ collects the set of offers/agents
that expressed their willingness to help $a$. By using some strategy,
$a$ selects one of the responding agents, say $b$.
The policy for choosing the responding agent can be programmed
 (e.g., by exploiting priorities, agent's knowledge on other agents,
random selection, trust criteria, utility and optimality considerations).
Once the choice has been made, $a$ establishes a rendez-vous with the
selected  agent
and
	\begin{list}{$\bullet$}{\topsep=1pt \parsep=0pt \itemsep=1pt}
	\item declares its availability to~$b$,
	\item  communicates the
 fulfillment of the request to the other agents.
 	\end{list}
The request and the obsolete offers are removed from the tuple space.
\end{list}
\item[(4)]
At that point in time, the transition can be considered completed for  the agent $a$.
By taking into account the information about the outcome of the coordination phase
in solving conflicts (point~(2)),
the agreement reached in handling requests (point~(3)), $a$
might need to modify its plan. If the replanning phase succeeds,
then $a$ will proceed with the execution of the next action in its local plan.
\end{enumerate}
Note that we provided separated descriptions for steps (3.a) and (3.b).
In a concrete implementation,
these two steps have to be executed in an interleaved manner,
to avoid that a fixed order
in sending requests and offers causes deadlocks or starvation.
Furthermore,  if an agent fails in executing an action,  then
it will skip the step~(3) and proceed with step~(4) in order to re-plan its activity.

\begin{figure}[htbp]
\centerline{\fbox{\begin{minipage}{0.85\textwidth}
\centerline{\includegraphics[width=0.83\textwidth]{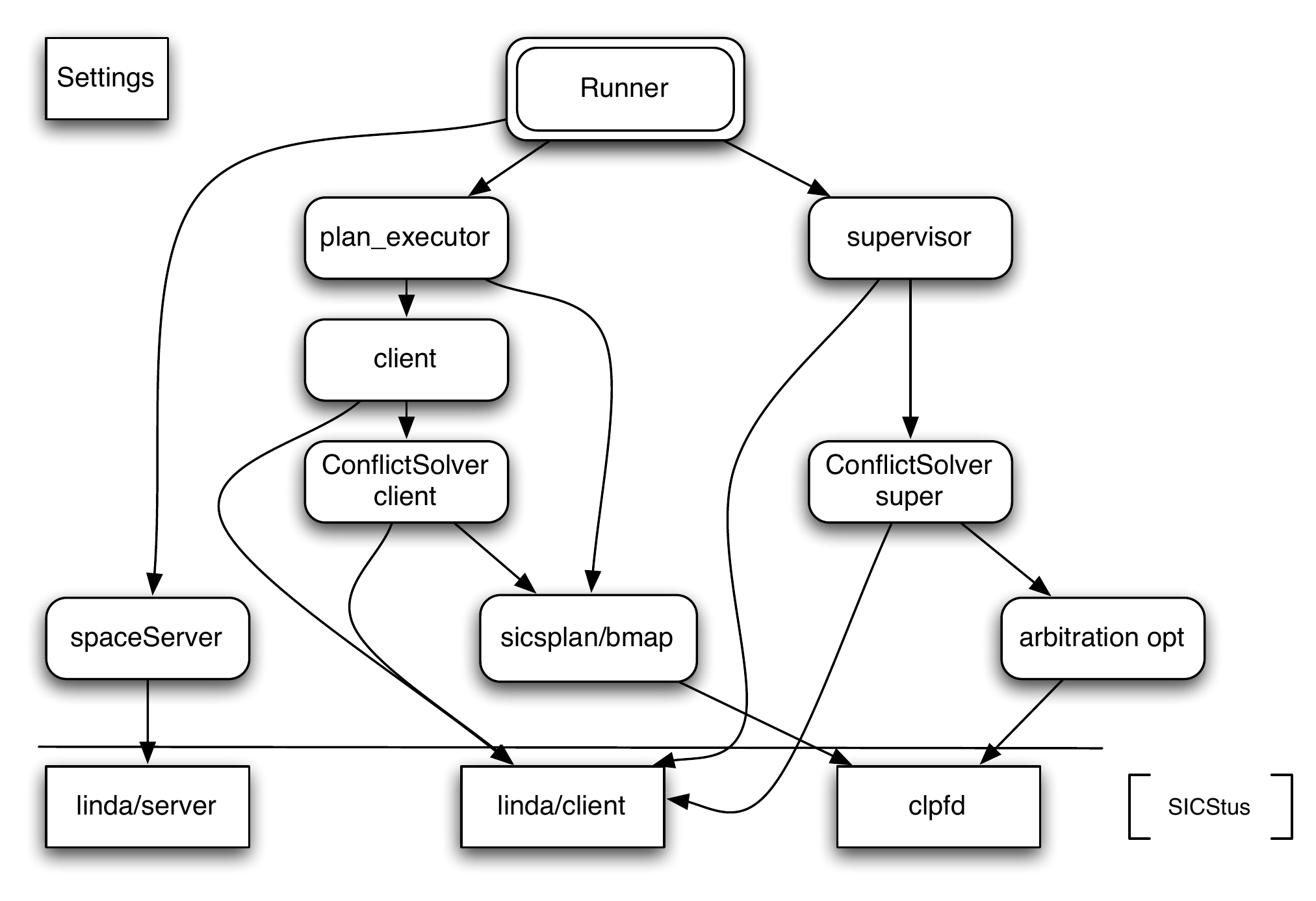}}
\end{minipage}}}
\caption{\label{architettura}The dependencies between modules in the system.
The modules' names recall the corresponding Prolog-files names.
The module \texttt{runner} is the starter of the application.
The module \texttt{settings} specifies user options (policies, strategies, etc.) and
the sources files containing the action descriptions, it is
imported by all the others (we omitted drawing the corresponding arcs, as well as the
nodes relative to less relevant SICStus libraries).}
\end{figure}

\subsection{The languages \BAAC, \BMAP, and \BMV}\label{sec:internal_comparison}

The language \BAAC, and its implementation, heavily relies on its foundations \BMAP\ and \BMV.
In this section we briefly compare these three languages to clarify which parts
of the solvers of the previous languages can be used for the implementation of \BAAC\ presented
in Subsection \ref{sec:implementation}.

Let us focus first on \BMV. This is a single agent framework. Therefore, considering a given action theory,
all fluents and actions are known to the single agent, and the language does not permit to specify private
fluents or actions.
Moreover,  \BMV\  allows one to specify static causal laws.
The syntax of fluent expressions and constraints is exactly the same as in \BAAC.
The syntax for executability and action effects is analogous to that of \BAAC. More precisely, in
\BMV, these laws take the forms:
\begin{itemize}
 \item \codetext{exectuable}($a$,$C$)
\item \codetext{causes}($x$,$C_1$,$C_2$),  where $C_1$ is the constraint that will hold in the next state if the action
$x$ is executed in a state where $C_2$ holds.
\end{itemize}
These are just syntactical variants of~(\ref{execDef}) and (\ref{dynaLaw}), respectively.
The semantics of \BMV\ is given via a transition system analogous to that
introduced for \BAAC. In particular, one might note that if a \BAAC\ action description involves a single agent
that knows all the fluents (and no communication laws are included), then
its semantics coincides with the one of the corresponding \BMV\ program obtained
by an immediat syntactical translation. The Prolog interpreter for \BMV\ is proved to be correct
and complete (for soundness the absence of static laws is needed, but this is the case of \BAAC, as presented here) with respect
to the semantics in~\cite{DFP-TPLP}.

Let us consider now \BMAP. It is a multiagent, centralized language, where
collective actions, namely actions that require more than one agent for being executed,
are allowed. For instance, a law of the form
$$
\mbox{ \codetext{action} $x$ \codetext{executable by} $a_1,a_2,\dots,a_n$ }
$$
specifies that agents $a_1,a_2,\dots,a_n$ may execute together the action $x$. In \BAAC, instead, in the domain
of an agent $a$, an action definition implicitly states that the action is executed by $a$
(hence, this is a particular case of the \BMAP\ law).
On the other hand, since the reasoner is centralized, conflicts among effects never occur and all
(concomitant) planned actions are always successfully executed.
The declaration of fluents in  \BMAP\ is analogous to that in \BAAC, whereas
\BMAP\ has a different syntax for dynamic laws, since they can refer directly to
action-occurrences. A \BMAP\ dynamic law has the form
$
\mbox{ \emph{Prec} \codetext{causes} \emph{Eff}}
$,
where \emph{Prec} and \emph{Eff} are constraints and at least one reference to an action \emph{x} must explicitly occur
in \emph{Prec}.
Such references are specified by exploiting action flags of the form \codetext{actocc($x$)}.

The semantics of \BMAP\ is given via the same notion of transition system used for \BMV\ and for \BAAC.
If a multi-agent action description in \BAAC, together with initial state and goal,
is such that during the plan, no conflict occurs, then the \BMAP\ action description obtained by
a simple (mostly one-to-one) translation, has exactly the same behaviour on the
transition system. Let us observe that in this translation, collective actions are not generated.

\subsection{Implementation issues}\label{sec:implementation}

A first prototype of the system has been  implemented in SICStus Prolog, using
the libraries \codetext{clpfd} for agents reasoning (by
 exploiting the interpreters for Action Description Languages
described in~\cite{DFP-POTSDAM,DFP-TPLP}), and the libraries
\codetext{system}, \codetext{linda/server},
and \codetext{linda/client} for handling process communication.

The system is structured in modules.
Figure~\ref{architettura} displays the modules composing the Prolog prototype
and their dependencies.
The modules \texttt{spaceServer} (via  \texttt{lindaServer}) and  \texttt{lindaClient}
implement the interfaces with the Linda tuple-space.
These modules support all the communications among agents.

Each autonomous agent corresponds to an instance of the module \texttt{plan\_executor},
which, in turn, relies on a planner (the module \texttt{sicsplan/bmap} in Figure~\ref{architettura})
for planning/re\-plan\-ning activities, and on
\texttt{client} for interacting with other agents in the system. As  explained previously,
a large part of the coordination is guided  by the module \texttt{supervisor}.
Notice that both the \texttt{supervisor} and \texttt{client} act
as Linda-clients.
Conflict resolution functionalities  are provided
to the modules \texttt{client} and \texttt{supervisor}
by the modules \texttt{ConflictSolver\_client}
and \texttt{ConflictSolver\_super},
respectively.
Finally, the \texttt{arbitration\_opt} module implements
the arbitration protocol(s). In the current code distribution, we provide an arbitration
strategy that maximizes the number of actions performed at each step.

Let us remark that all the policies exploited in  coordination, arbitration, and conflict handling
can be customized by simply providing a different implementation of
individual  predicates exported by
the corresponding modules. For instance, to implement a conflict resolution strategy different
from the round-robin described earlier, it suffices to add to the system a
new implementation of  the module \texttt{ConflictSolver\_super}
(and for \texttt{ConflictSolver\_client}, if the specific strategy requires
an active role of the conflicting agents).
Similar extensions can be done for \texttt{arbitration\_opt}.

The system execution is rooted in the server process {\tt runner}---written
either for Linux ({\tt .sh}) or for Windows ({\tt .bat})
platforms, in
charge of generating the
connection address that must be used by the client processes.

The file {\tt settings.pl} describes the planning problem to be solved.
In particular, the user must specify in this file, through Prolog facts,
the number and the names of these files containing the action descriptions,
a bound on the maximum length of the plan, and
the selected strategies for conflict resolution and arbitration
(default choices can be used).

As far as the reasoning/planning module is concerned, we slightly
modified the interpreters of the $B^{MV}$ and the $B^{\sf MAP}$ languages
\cite{DFP-POTSDAM,DFP-TPLP} to accept the extended
syntax presented here.
However, the system is open to further extensions and different planners
(even not necessarily based on Prolog technology)
can be easily integrated thanks to the simple interface with the module
\texttt{plan\_executor}, which consists of a few Prolog predicates.

Currently, two planners have been integrated in the system:
\texttt{sicsplan} is the constraint logic programming planner for the
single-agent action language $B^{MV}$; \texttt{bmap} is instead
a constraint logic programming engine that supports centralized
planning for multi-agent systems (capable, e.g., of collaborating
in pursuing a common goal). Thus, the implementation allows each
individual agent (according to the discussion from the previous sections)
to be itself a complex system composed of multiple agents (operating
in a cooperative fashion and planning in a centralized manner).

To accommodate for this perspective, the design of
the supervisor has been modified.
The framework allows each concurrent planner that
executes a multiple-action step, to specify the desired granularity of the conflict resolution phase.
This is done by specifying  (for each step in a plan) a partition of the
set of actions composing the step into those subsets of actions that
have to be considered independently and as a whole.

For instance,
in the next section we describe a  specification
of a coordination problem between two multi-agent systems.
Each multi-agent system develops a plan in a centralized manner.
Each step of such plans
consists of a set of, possibly complex, actions (instead of a single action,
as happens for the planner \texttt{sicsplan}).
The
conflicts between the multi-agent plans occurring during the $(i)$-th state transition
are identified/resolved by considering a
single action of each $(i)$-th step proposed by each planner.

Let us make some considerations about the soundness of the implementation. Let
us consider one step $i+1$ in the construction of the trajectory. The state sequence
already constructed is $\vec{v} = \langle v_0, \dots, v_i\rangle$. The agents propose some
actions for execution; the overall set of all actions proposed by all agents is
$Y_{i+1} = \{y_1, \dots, y_k\}$. Agents propose for execution actions that are
executable in $\vec{v}$. At the implementation level, the soundness property is guaranteed by
the correctness of the \texttt{sicsplan/bmap} module---see Section \ref{sec:internal_comparison}.

Let us denote with $C(y_j)$ the constraint that captures the
 effects of action $y_j$; i.e., if the action $y_j$ has dynamic causal laws
 $y_j \codetext{ causes } E_r \codetext{ if } P_r$ for $r=1,\dots, m$, then
 $$C(y_j) \equiv \bigwedge_{r=1}^m P_r \rightarrow E_r.$$
 Let $A(y_j)$ be a Boolean variable, intuitively denoting whether the supervisor has
 selected action $y_j$ for execution at time $i+1$.

The \texttt{arbitration\_opt} implements an arbitration protocol $\Phi(\vec{v},Y_{i+1})$
producing a substitution for $\{A(y_1), \dots, A(y_k)\}$ such that the constraint
$$\bigwedge_{j=1}^k \Phi(\vec{v},Y_{i+1})(A(y_j)) \rightarrow C(y_j)$$
has a $\vec{v}$-solution $\sigma$.

For example, in the current code distribution, the protocol $\Phi$ is defined as a
substitution that maximizes $\sum_{j=1}^k A(y_j)$.

From these definitions and from the properties of
\texttt{sicsplan/bmap}, we have that
$\langle \vec{v}, \{y_j\:|\: j \in \{1,\dots,k\},  \Phi(\vec{v},Y_{i+1})(y_j)=1\}, \ine(\sigma,\vec{v}) \rangle$
is a valid state transition.

If the conflict resolution is left to the agents, then the protocol $\Phi$ is the outcome of the
conflict resolution procedure, e.g., the round-robin analysis of the conflicting actions
described in Section~\ref{sec:inconsistency}, which is currently implemented.
It is immediate to check that the round-robin procedure produces a protocol $\Phi$ that
satisfies the properties shown above.

Due to the generality of the language for agent-based on-conflict resolution, the correctness
of any conflict resolution procedure must be independently proved. Correctness is not an immediate consequence
of the language itself but is dependent on the specific on-conflict declaration are used in the specific procedure.

\subsection{The volleyball domain}\label{volleysect}
Let us describe a specification in  \name\
of a coordination problem between two multi-agent systems---an
extension of the   domains described
in Examples \ref{volley1}--\ref{volley4}.
There are two teams: \emph{black} and
\emph{white} whose objective is to score a point, i.e., to
throw the ball in the field of the other team (passing over the net)
in such a way that no player of the other team can reach the ball
before it touches the ground. Each team is
modeled as a multi-agent system
that elaborates its own plan in a centralized manner (thus,
each step in the plan consists of a set of actions).

\newsavebox{\verbbox}

\begin{figure}[htbp]
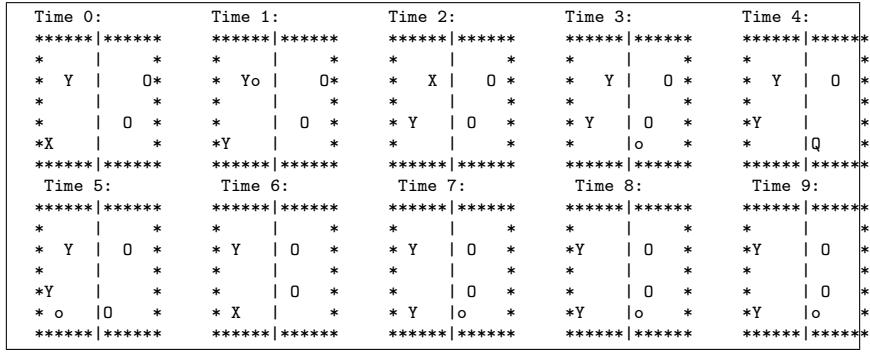

{\scriptsize
\centering
\begin{lrbox}{\verbbox}
\begin{minipage}{.917\textwidth}
\begin{verbatim}
  Time 0:           Time 1:           Time 2:           Time 3:           Time 4:
  ******|******     ******|******     ******|******     ******|******     ******|******
  *     |     *     *     |     *     *     |     *     *     |     *     *     |     *
  *  Y  |    O*     *  Yo |    O*     *   X |   O *     *   Y |   O *     *  Y  |  O  *
  *     |     *     *     |     *     *     |     *     *     |     *     *     |     *
  *     |  O  *     *     |  O  *     * Y   | O   *     * Y   | O   *     *Y    |     *
  *X    |     *     *Y    |     *     *     |     *     *     |o    *     *     |Q    *
  ******|******     ******|******     ******|******     ******|******     ******|******
   Time 5:           Time 6:           Time 7:           Time 8:           Time 9:
  ******|******     ******|******     ******|******     ******|******     ******|******
  *     |     *     *     |     *     *     |     *     *     |     *     *     |     *
  *  Y  |  O  *     * Y   | O   *     * Y   | O   *     *Y    | O   *     *Y    | O   *
  *     |     *     *     |     *     *     |     *     *     |     *     *     |     *
  *Y    |     *     *     | O   *     *     | O   *     *     | O   *     *     | O   *
  * o   |O    *     * X   |     *     * Y   |o    *     *Y    |o    *     *Y    |o    *
  ******|******     ******|******     ******|******     ******|******     ******|******
\end{verbatim}
\end{minipage}
\end{lrbox}\centerline{\fbox{\usebox{\verbbox}}}
\caption{\label{Fig:volley}A representation of an execution of the volleyball domain
}
}
\end{figure}

The playing field is discretized by fixing a $\codetext{linex}
\times \codetext{liney}$
rectangular grid that determines the positions where the players
(and the ball) can move (see Fig.~\ref{Fig:volley}).
The leftmost (rightmost) cells are those of the black (white) team,
while the net ($x=6$) separates the two subfields.
There are $p$ players per team ($p=2$ in Fig.~\ref{Fig:volley})---concretely,
the fact \codetext{num(2)} is added to the theory.
The
allowable actions are:
\codetext{move}$(d)$,
\codetext{throw}$(d,f)$,
and \codetext{whistle}.
During the defense time,
the players can move 
to catch the ball and/or to re-position themselves on the court.
When a player reaches the ball
(s)he will have the ball and will throw
the ball again.
A team scores a point either if it throws
the ball to a cell in the opposite subfield
that is not reached by any player of the other team in the defense time,
or if the
opposite team throws the ball
in the net. The captain (first player) of each team is in charge of checking
if a point has been scored. In this case, (s)he \codetext{whistle}s.


Each team (either {\bf black} or {\bf white}) is modeled as a centralized multi-agent system, which
acts as a singe agent in the interaction with the other team.
Alternative options in modeling are also possible---for instance,
one could model each single player as an independent agent that
develops its own plan and interacts with all other players.
The two teams have the goal of scoring a  point:
\codetext{goal(point(black) eq 1).} for blacks and
\codetext{goal(point(white) eq 1).} for whites.

At the beginning
of the execution every team has a winning strategy, developed as a local plan;
these
are possibly revised after each play to accommodate for the new
state of the world reached.  An execution (as printed by the system)
is reported in Fig.~\ref{Fig:volley}, for
a plan length of $9$. The symbol {\tt 0} (respectively, {\tt Y}) denotes the white (respectively, black) players,
{\tt Q} (resp. {\tt X}) denotes a white player with the ball.
The throw moves applied are:

{\begin{tabular}{lclc}
{\footnotesize \tt [player(black,1)]:throw(ne,3)} & (time 1) &
{\footnotesize \tt [player(black,2)]:throw(se,3)} & (time 3) \\
{\footnotesize \tt [player(white,1)]:throw(w,5)} & (time 5) &
{\footnotesize \tt [player(black,1)]:throw(e,5)} & (time 7)
\end{tabular}}

\noindent Let us observe that, although it would be in principle possible for the
white team to reach the ball and throw it within the time
allowed, it would be impossible to score a point. Therefore,
players prefer to avoid to perform any move.

\smallskip

The complete description of the encoding of this domain is available at
 {\small\url{http://www.dimi.uniud.it/dovier/BAAC}}. The repository includes
 also additional domains---e.g., a domain inspired by games
involving one ball and two-goals, as found in
soccer.
Although the encoding might seem similar to that
of volleyball, the possibility of contact between two players 
makes this encoding  more complex.
Indeed,
thanks to the fact that  the net separates the two teams,
in the volleyball domain rules like the following one suffice to avoid collisions:

\codetext{\footnotesize
\begin{tabular}{l}
always(pair(x(A),y(A)) neq pair(x(B),y(B))) :-\su\\
\tab\tab\tab    A=player(black,N),B=player(black,M), num(N), num(M), N<M.\giu
\end{tabular}
}\par

\noindent
In a soccer world this is not true because only the
supervisor can be aware, in advance, of possible contacts
between different team players originating from concurrent actions.
This generates
interesting concurrency problems, e.g., concerning the
ball possession after a contact.
A simple way to address this problem consists in assigning a fluent to each field cell,
whose value can be $-1$ (free), $0$ (resp., $1$) if a white (resp. black) player is in the cell.
The supervisor identifies a conflict when two opponent players move to the same cell,
thus assigning to that fluent a different value.
In this case, the supervisor  arbitrarily enables one action,
the other agent waits a turn to retry the action:

\codetext{\footnotesize
\begin{tabular}{l}
action  act([A],move(D))  on\_failure retry\_after 1 on\_conflict arbitrate :- \su \\
\tab\tab\tab  agent(A), direction(D).\giu
\end{tabular}
}

\section{Conclusions and future work}\label{sec:conclusions}

In this paper, we illustrated the design of a high-level
action description language for the description of multi-agent domains. The
language enables the description of agents with individual goals operating
in a shared environment. The agents can  explicitly interact (by requesting
help from other agents in achieving their own goals) and implicitly cooperate
in resolving conflicts that may arise during execution of their individual
plans. The main features of the framework we described in this paper
have been realized into an implementation, based on
SICStus Prolog.
The implementation is fully distributed, and uses Linda
to enable communication among agents.
Such a prototype is currently being refined and extended with further features.

There have been many agent programming languages  such as
the BDI agent programming AgentSpeak \cite{agentspeak},
(as implemented in Jason \cite{jason}),
JADE \cite{jade} (and its extension Jadex \cite{jadex}), ConGolog \cite{congolog}, IMPACT
\cite{impact}, 3APL~\cite{3apl}, GOAL~\cite{goal}. A good
comparison of many of these languages can be found in \cite{survey}.
The emphasis of the effort presented in this paper is to expand
our original work on constraint-based modeling of agents based
on action languages. The generalization to a constraint-based
multi-agent action language has been presented in \cite{DFP-POTSDAM}. In this
paper we demonstrate a further extension to encompass distributed
reasoning and distributed planning. Thus, the focus of the proposal
remains on the level of creating an action language and demonstrating
the suitability of constraint-based technology to support it. As such,
we do not propose here a new agent programming language, rather we
push an action language perspective and how action languages scale to multi-agent
domains; our work could be used as the underlying formalism for the
development of new agent programming languages. In this sense, our
proposal is different than many of the MAS development platforms, which
focus  on programming languages for MAS and on complex protocols for
advertising and interaction among agents (e.g., FIPA).

The choice of Linda came about for simplicity; we required the use
of a CLP platform and SICStus provides support for both Linda and constraint
handling---as few other distributed communication platforms
 (e.g., OAA \cite{oaa}). In the long term, we envision mapping
our agent design on a MAS infrastructure that enables discovery and addition
of agents,  handles network-wide distribution of agents, mapping the
exchange of constraints to a standard agent communication language
(e.g., FIPA-ACL/FIPA-SL \cite{fipa}). This will require a non-trivial engineering work,
to map the reasoning with action languages (e.g., planning) to a platform that
is not constraint-based---we are currently exploring the problem in the context
of Jason \cite{jason}.

The work is an initial proposal that already shows strong potential and
several avenues of research. The immediate goal in the improvement
of the system consists of adding refined strategies and coordination
 mechanisms, involving for instance, payoff, trust, etc.
Then, we intend to evaluate the performance and quality
of the system in several multi-agent domains (e.g.,
game playing scenarios, modeling of auctions, and other
domains requiring distributed planning).
We also plan to investigate strategies to enhance performance
by exploiting features provided by the constraint solving libraries
of SICStus (e.g., the use of the table constraint~\cite{Bartak08}).

We will investigate the use of future references in the fluent
constraints (as fully supported in $B^{MV}$)---we believe this feature
may provide a more elegant approach to handle the requests
among agents, and it is necessary to enable the expression of complex
interactions among agents (e.g., to model forms of negotiation with
temporal references). In particular, we view
this platform as ideal to experiment with models of \emph{negotiation}
(e.g., as discussed in \cite{chiaki}) and to deal with
\emph{commitments} \cite{commit} (which often require temporal references).

We will also explore the implementation of different strategies
associated to conflict resolution; in particular, we are interested
in investigating how to capture the notion of ``trust'' among agents,
as a dynamic property that changes depending on how reliable agents
have been in providing services to other agents (e.g., accepting to
provide a property but failing to make it happen).
Also concerning trust evaluation, different approaches can be integrated in the system.
For instance, a ``controlling entity'' (e.g., either the supervisor or a privileged/elected
agent) could be in charge of assigning the ``degree of trust'' of each agent.
Alternatively, each single agent could develop its own opinion on other
agents' reliability, depending on the behavior they manifested in past interactions.

Finally, work is  needed to expand the framework to enable greater
flexibility in several aspects, such as:
\begin{itemize}
\item
Allow  deadlines for requests---e.g., by allowing axioms of the
form\\
\centerline{\codetext{request } $C_1$ \codetext{ if } $C_2$ \codetext{until } $T$}
indicating that the request is valid only if accomplished within
$T$ time steps.
\item
Allow  constraint based delays for requests:\\
\centerline{\codetext{request } $C_1$ \codetext{ if } $C_2$ \codetext{while } $C_3$}
indicating that the request is still valid while constraint $C_3$ is entailed.
\item
Allow dynamic changes in the agents' knowledge about other agents
(e.g., an action might make an agent aware of the existence of other agents),
or about the world
(e.g., an action might change the rights another agent has to access/modify some fluents).
\end{itemize}

\paragraph*{Acknowledgments}
The authors wish to thank the anonymous reviewers for their insightful comments.

\bibliographystyle{plain}

\begin{thebibliography}{}

\bibitem{Bartak08}
{\sc Bart{\'a}k, R.} {\sc and} {\sc Toropila, D.} 2008.
\newblock Reformulating constraint models for classical planning.
\newblock In {\em  Int. Florida AI
 Research Society Conference}, AAAI Press, 525--530.

\bibitem{jade}
{\sc Bellifemine, F.}, {\sc Caire, G.}, {\sc and} {\sc Greenwood, D.} 2007.
\newblock {\em {Developing Multi-Agent Systems with JADE}}.
\newblock John Wiley \& Sons.

\bibitem{jason}
{\sc Bordini, R.}, {\sc H{\"u}bner, J.}, {\sc and} {\sc Wooldridge, M.} 2007.
\newblock {\em Programming Multi-agent Systems in {AgentSpeak} using {Jason}}.
\newblock J. Wiley and Sons.

\bibitem{jadex}
{\sc Braubach, L.}, {\sc Pokahr, A.}, {\sc and} {\sc Lamersdorf, W.} 2005.
\newblock {Jadex: a BDI-Agent System Combining Middle-ware and Reasoning}.
\newblock In {\em Software Agent-based Applications, Platforms and Development
  Kits}. Springer Verlag.


\bibitem{linda}
{\sc Carriero, N.}  {\sc and} {\sc Gelernter, D.} 1989.
\newblock {Coordination Languages and their Significance}.
\newblock {\em Communications of the ACM\/}~{\em 32\/}~4.

\bibitem{oaa}
{\sc Cheyer, A.} {\sc and} {\sc Martin, D.} 2001.
\newblock {The Open Agent Architecture}.
\newblock {\em Journal of Autonomous Agents and Multi-Agent Systems\/}~{\em
  4,\/}~1, 143--148.

\bibitem{3apl}
{\sc Dastani, M.}, {\sc Dignum, F.}, {\sc and} {\sc Meyer, J.-J.} 2003.
\newblock {3APL}: A programming language for cognitive agents.
\newblock {\em ERCIM News\/}~{\em 53}, 28--29.

\bibitem{goal}
{\sc {de Boer}, F.}, {\sc Hindriks, K.}, {\sc {van der Hoek}, W.}, {\sc and}
  {\sc Meyer, J.} 2005.
\newblock {A Verification Framework for Agent Programming with Declarative
  Goals}.
\newblock {\em JAL,\/}~{\em 5}, 277--302.

\bibitem{congolog}
{\sc {De Giacomo}, G.}, {\sc Lesp{\`e}rance, Y.}, {\sc and} {\sc Levesque, H.}
  2000.
\newblock {ConGolog}, a concurrent programming language based on the situation
  calculus.
\newblock {\em AIJ,\/}~{\em 121,\/}~1--2, 109--169.

\bibitem{DFP-POTSDAM}
{\sc Dovier, A.}, {\sc Formisano, A.}, {\sc and} {\sc Pontelli, E.} 2009.
\newblock Representing multi-agent planning in {CLP}.
\newblock In {\em LPNMR },
  Lecture Notes in Computer Science, vol. 5753. Springer, 423--429.

\bibitem{DFP-TPLP}
{\sc Dovier, A.}, {\sc Formisano, A.}, {\sc and} {\sc Pontelli, E.} 2010.
\newblock Multivalued action languages with constraints in {CLP(FD)}.
\newblock {\em Theory and Practice of Logic Programming\/}~{\em 10,\/}~2,
  167--235.

\bibitem{fagin95}
{\sc Fagin, R.} et al.  1995.
\newblock {\em Reasoning about knowledge}.
\newblock The MIT Press.


\bibitem{GL98}
{\sc Gelfond, M.} {\sc and} {\sc Lifschitz, V.} 1998.
\newblock Action languages.
\newblock {\em Electronic Transactions on Artificial Intelligence\/}~{\em 2},
  193--210.

\bibitem{Gerbrandy06}
{\sc Gerbrandy, J.} 2006.
\newblock Logics of propositional control.
\newblock In \cite{DBLP:conf/atal/2006}, 193--200.

\bibitem{fipa}
{\sc Hayzelden, A.} {\sc and} {\sc Bourne, R.} 2001.
\newblock {\em Agent Technology for Communication Infrastructures}.
\newblock John Wiley \& Sons.

\bibitem{commit}
{\sc Mallya, A.} {\sc and} {\sc Huhns, M.} 2003.
\newblock Commitments among agents.
\newblock {\em IEEE Internet Computing\/}~{\em 7,\/}~4, 90--93.

\bibitem{survey}
{\sc Mascardi, V.}, {\sc Martelli, M.}, {\sc and} {\sc Sterling, L.} 2004.
\newblock Logic-based specification languages for intelligent agents.
\newblock {\em Theory and Practice of Logic Programming\/}~{\em 4,\/}~4,
  495--537.

\bibitem{DBLP:conf/atal/2006}
{\sc Nakashima, H.}, {\sc Wellman, M.~P.}, {\sc Weiss, G.}, {\sc and} {\sc
  Stone, P.}, Eds. 2006.
\newblock {\em International Joint Conference on
  Autonomous Agents and Multiagent Systems}. ACM.

\bibitem{agentspeak}
{\sc Rao, A.} 1996.
\newblock {AgentSpeak: BDI Agents Speak Out in a Logical Computable Language}.
\newblock In {\em European Workshop on Modeling Autonomous Agents in a
  Multi-Agent World}.

\bibitem{centralized2}
{\sc Sauro, L.}, {\sc Gerbrandy, J.}, {\sc van~der Hoek, W.}, {\sc and} {\sc
  Wooldridge, M.} 2006.
\newblock Reasoning about action and cooperation.
\newblock See \cite{DBLP:conf/atal/2006}, 185--192.

\bibitem{chiaki}
{\sc Son, T.}, {\sc Pontelli, E.}, {\sc and} {\sc Sakama, C.} 2009.
\newblock Logic programming for multiagent planning with negotiation.
\newblock In {\em Int. Conference on Logic Programming}. Springer,
  99--114.

\bibitem{SpaanGV06}
{\sc Spaan, M. T.~J.}, {\sc Gordon, G.~J.}, {\sc and} {\sc Vlassis, N.~A.}
  2006.
\newblock Decentralized planning under uncertainty for teams of communicating
  agents.
\newblock  In \emph{AAMAS}, ACM Press, 249--256.

\bibitem{impact}
{\sc Subrahmanian, V.~S.}, {\sc Bonatti, P.}, {\sc Dix, J.}, {\sc Eiter, T.},
  {\sc Kraus, S.}, {\sc Ozcan, F.}, {\sc and} {\sc Ross, R.} 2000.
\newblock {\em Heterogeneous Agent Systems: Theory and Implementation}.
\newblock  MIT Press.

\bibitem{HoekJW05}
{\sc van~der Hoek, W.}, {\sc Jamroga, W.}, {\sc and} {\sc Wooldridge, M.} 2005.
\newblock A logic for strategic reasoning.
\newblock In {\em AAMAS}, ACM Press, 157--164.

\end{thebibliography}


\end{document}